\documentclass[aps,prb,twocolumn,showpacs,longbibliography]{revtex4-1}
\usepackage{bm}
\usepackage{color} 
\usepackage{graphicx}
\usepackage{epsfig}
\usepackage{hyperref}
\usepackage{natbib}
\usepackage{amsmath}
\usepackage{amssymb}

\newcommand{\rmd}{{\rm d}}
\newcommand {\e}{{\rm e}}
\newcommand {\Tr}{{\rm Tr}\,}
\renewcommand {\i}{{\rm i}}
\renewcommand {\Re}{\mathop{\mathrm{Re}}\nolimits}

\newcommand{\av}[1]{\left\langle #1\right\rangle}
\newcommand{\ket}[1]{\left| #1\right\rangle}

\newcommand{\Gc}{\Gamma_{\rm C}}
\newcommand{\Gx}{\Gamma_{\rm X}}
\newcommand{\Nc}{N_{\rm C}}
\newcommand{\Nx}{N_{\rm X}}

\begin{document}

\title{Time-resolved second-order correlations of microcavity photons}

\author{A.\,V.\,Poshakinskiy}\email{poshakinskiy@mail.ioffe.ru}
\author{A.\,N.\,Poddubny}
\affiliation{Ioffe Physical-Technical Institute, Russian Academy of Sciences, 194021 St.~Petersburg, Russia}
\begin{abstract}
The time dependence of the correlations between the photons, emitted from the microcavity with embedded quantum dot under incoherent pumping, is studied theoretically. Analytical expressions for the second-order correlation function $g^{(2)}(t)$ are presented in strong and weak coupling regimes. At moderate pumping the correlation function demonstrates Rabi oscillations, while at larger pumping it shows monoexponential decay. The decay time of the correlations nonmonotonously depends on the pumping value and has a maximum corresponding to the self-quenching transition.
\end{abstract}

\maketitle

\section{Introduction}
Semiconductor quantum dots form a promising platform for quantum optics devices, including single photon emitters and emitters of entangled photon pairs.
\cite{kavbamalas,muller2009,Dousse2010,Kuhn2010}
 The quantum dot-based light sources can be characterized by means of photon-photon correlation spectroscopy, i.e. by measuring the second-order correlation function $g^{(2)}(t)$ between two photons with the delay $t$.\cite{Carmichael}
Multiple experimental observations of the antibunching [$g^{(2)}(0)<1$] of the photons  emitted from the quantum dots are already available.\cite{michler2000,Yuan2002,calic2011,Abbarchi2012}
  One of the possible routes 
 to further enhance  the performance of these  light  sources
is to resonantly couple the  quantum dot exciton   with the photonic mode, confined inside the  microcavity in all three spatial directions.\cite{Dousse2010} 
The physics of such quantum  microcavites becomes especially rich in the strong coupling regime, where the new quasiparticles, exciton polaritons, are formed due to the interaction between the excitons and the cavity photons.\cite{kavbamalas,Khitrova2006,Reitzenstein2010,Dousse2010,nomura2010,yanda2011}

Here, we study the time dependence of the second-order correlations between the photons, emitted from the quantum dot microcavity.
We analyze the case of incoherent nonresonant pumping of the excitons inside the quantum dot. This is a specific feature of the considered
problem, different from  the resonant fluorescence scheme, generally employed  in atomic cavities.\cite{kimble1977,Walls1979,Hennrich2005,Jabri2011}
Comprehensive studies of the stationary  function $g^{(2)}(0)$ have been carried out in Refs.~\onlinecite{laussy2009,Valle2011}.
It has been demonstrated, that the value of $g^{(2)}(0)$  increases with pumping strength and exhibits qualitatively different behavior in weak and strong coupling regimes.
 Hence, we focus our attention on the temporal dynamics of the  correlations.
 Our main goal is to derive transparent analytical answers for the correlation function $g^{(2)}(t)$ as function of pumping both in strong and in weak coupling regimes.
We show, how the lifetime of the polariton eigenstates determines the correlation decay rate, and how the spacings between the polariton energy levels are manifested in  the Rabi oscillations.
We consider the  stationary pumping case, favorable for studies of 
the time dependence of the correlations.
Experimentally, such regime can be realized in  
 quantum dot microcavities driven by electrical pumping \cite{Schneider2012} 
or continuous optical pumping.
This situation should be distinguished from the case of 
 pulsed pumping,\cite{poddubny2012} where the time dependence of the signal depends on the individual pulse shape.\cite{Kessler2012,Illes2010}

The rest of the paper is organized as follows.  In Sec.~\ref{sec:model} the model and the calculation approach are described.  Sec.~\ref{sec:strong} and Sec.~\ref{sec:weak} present the theory developed in strong and weak coupling regimes, respectively. Paper results are summarized in Sec.~\ref{sec:summary}. Auxiliary derivations are given in Appendices~\ref{ApS} and \ref{ApQ}.
\section{Model}\label{sec:model}
We consider zero-dimensional microcavity 
where the  single photon mode is coupled to the single excitonic state of the quantum dot. Polarization degrees of freedom of both photons and excitons are disregarded for simplicity. Under these assumptions the  Hamiltonian of the studied system has the standard form\cite{kavbamalas}
\begin{equation}\label{eq:ham}
H=\hbar\omega_0 c^\dag c + \hbar \omega_0 b^\dag b + \hbar g (c^\dag b + c b^\dag) \:.
\end{equation}
Here, $\omega_0$ is the resonance frequency of the cavity, tuned to the   exciton resonance, $c$ and $c^\dag$ are the bosonic annihilation and creation operators for cavity mode ($[c,c^{\dag}]=1$), $b$ and $b^\dag$ are corresponding ``fermionic'' operators for the single exciton mode ($b^2=0$, $\{b,b^\dag\}=1$), and $g$ is the light-exciton coupling constant. Eq.~\eqref{eq:ham} corresponds to the  quantum dot smaller than the exciton Bohr radius. To consider the case of large quantum dot one should generalize the model of Refs.~\onlinecite{JETP2009, Poddubny2013}.
\begin{figure*}[tb]
\includegraphics[width=0.95\textwidth]{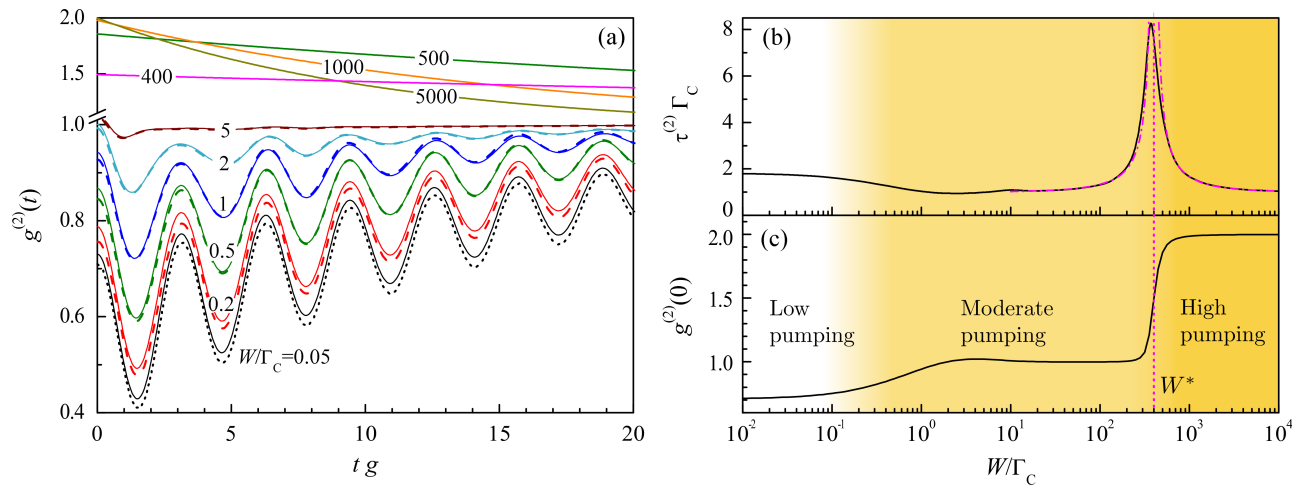}
\caption{(Color online) (a) Time dependence of the correlation function $g^{(2)}(t)$ in the strong coupling regime, $g/\Gc=10$. Curves are plotted for $\Gx=0.1\Gc$ and various pumping rates $W/\Gc$ shown in graph. Solid lines are obtained numerically. Dotted line corresponds to the low pumping regime and is plotted after Eq.~(\ref{eq:g2t_strong_linear}). Dashed lines present analytical results for the case of moderate pumping and are plotted after Eq.~(\ref{eq:g2t_strong}). 
Panels (b) and (c) show  the correlation function  decay time calculated after Eq.~\eqref{eq:decay_time} and the stationary correlator $g^{(2)}(0)$, respectively, as a function of the pumping. The titles correspond to the regimes analyzed in Sec.~\ref{sec:low}--\ref{sec:high}. The dotted vertical line indicates the value of the critical pumping $W^*=4g^2/\Gc=400~\Gc$, where the sharp peak of correlations lifetime $\tau^{(2)}$ occurs. The asymptotics of the time $\tau^{(2)}$ near the critical point plotted after Eq.~\eqref{Qtaug2} is shown by dash-dotted curve.
}
\label{fig:strong}
\end{figure*}

In order to determine  the intensity of emission  from the cavity one should  also  introduce the processes of particles generation and decay.  We consider  incoherent continuous pumping of excitons into the quantum dot with the rate $W$. More detailed discussion of the pumping mechanism can be found in Ref.~\onlinecite{JETP2009}. Excitonic mode is characterized with the non-radiative damping $\Gx$.
 Photons can escape the cavity through the mirrors with the rate $\Gc$. Hence, the full system state is described by the density matrix $\rho$ and its evolution is determined by the equation $\rmd\rho/\rmd t= \mathcal{L}[\rho]$
with the  Lindblad terms in the Liouvillian $\mathcal{L}$, accounting for damping and pumping:\cite{kavbamalas}
\begin{align}\label{liou}
&\mathcal{L}[\rho] = -\frac{\i}{\hbar}[H,\rho ] + \frac{\Gc}{2}(2c \rho c^\dag - c^\dag c \rho - \rho c^\dag c ) \\
& + \frac{\Gx}{2}(2b \rho b^\dag - b^\dag b \rho - \rho b^\dag b )
+ \frac{W}{2}(2b^\dag \rho b - b b^\dag \rho - \rho b b^\dag ) \nonumber \:.
\end{align}
Stationary  density matrix $\rho_0$  satisfies the equation $\mathcal{L}[\rho_0]=0$. One can calculate the number of photons in cavity $\Nc =\av{c^\dag c}$ and the exciton occupation number $\Nx =\av{b^\dag b}$  as
\begin{equation}\label{eq:NcNx_def}
\Nc = \Tr (c^\dag c \rho_0)\:, \quad \Nx = \Tr (b^\dag b \rho_0)\:,
\end{equation}
where $\Tr$ stands for the operator trace and angular brackets denote the quantum mechanical expectation value. Detailed study of the dependence of these numbers on the pumping and on the other parameters can be found in Refs.~\onlinecite{laussy2009,Poddubny2010prb}. 
The goal of this work is to analyze the time dependence of the fluctuations of the emission intensity from the cavity. They  are described by the  correlator $g^{(2)}(t)$ determining the probability to register two photons with the time delay $t$:\cite{Carmichael}
\begin{equation}
g^{(2)}(t)=\frac{1}{\Nc^2}\, \av{c^\dag(0) c^\dag(t) c(t) c(0)} \:.\label{eq:g2_def0}
\end{equation}
Eq.~\eqref{eq:g2_def0} presents the simplest definition of the correlation function, suitable for the following analytical treatment. The more general expression, taking into account the finite response rate and spectral window of the photon detector for two- and multiple- photon correlations is given in Ref.~\onlinecite{valle2012}.
The way of calculation of $g^{(2)}(t)$ is provided by the quantum regression theorem,\cite{Carmichael}
\begin{equation}\label{eq:g2_def}
g^{(2)}(t)=\frac{1}{\Nc^2}\, \Tr [ c^\dag c \chi(t)] \:,
\end{equation}
where the evolution of the operator $\chi(t)\equiv {\rm e}^{\mathcal L t}[c \rho_0c^\dag]$ is governed by the dynamic equation
\begin{equation}\label{eq:g2_eq}
\frac{d\chi}{dt} = \mathcal{L}[\chi] \:, \quad \chi(0) = c \rho_0 c^\dag \:.
\end{equation}
For zero time delay Eq.~\eqref{eq:g2_def} assumes the form
\begin{equation}\label{eq:g20_def}
g^{(2)}(0)=\frac{1}{\Nc^2}\, \Tr (c^\dag c^\dag c c \rho_0)\:.
\end{equation}
For large time delays the correlator tends to unity, 
$g^{(2)}(t\to\infty)=1$,
 because the probabilities of detection of two photons become independent.
%
\section{Strong coupling regime}\label{sec:strong} 

In this section we  analyze the time dependence $g^{(2)}(t)$ in the strong coupling regime, when $g \gg \Gc, \Gx$.
We first present a general overview of the results and then provide a detailed analytical description in different regimes, determined by the strength of pumping.

Our main calculation results are summarized in Fig.~\ref{fig:strong}. Panel (a) shows the dynamics of the correlator $g^{(2)}(t)$, while panels (b) and (c) present the average lifetime of the correlations
\begin{equation}
 \tau^{(2)}=\frac{\int_0^\infty {\rm d}t  [g^{(2)}(t)-1]t}{\int_0^\infty {\rm d}t [g^{(2)}(t)-1]}\label{eq:decay_time}
\end{equation}
and the stationary value $g^{(2)}(0)$, respectively. Fig.~\ref{fig:strong} demonstrates, that the dependence of the time-resolved correlations on pumping is not trivial. In Sec.~\ref{sec:low}--~\ref{sec:high} the following qualitatively different regimes are 
described:

(A) {\it Low pumping}, $W\ll \Gc$. In this case the correlation function $g^{(2)}(t)$ is less than unity at $t=0$ (antibunching) and demonstrates Rabi oscillations with the frequency $2g$. The decay rate of the oscillations  is equal to the average of the exciton and photon decay rates.

(B) {\it Moderate pumping}, $\Gc \lesssim W \ll g^2/\Gc$.
Growth of the pumping intensity leads to the decrease of both the period and the lifetime of the oscillations. 
In a wide range of higher pumping intensities $\Gc \ll W \ll g^2/\Gc $ the emission statistics is Gaussian and the correlation function is close to unity and almost time-independent. This can be understood as a lasing regime for the dot, strongly coupled to the cavity mode.

(C) {\it High pumping}. The pumping value $W^*= 4g^2/\Gc$ corresponds to the transition from the lasing regime to the so-called self-quenching regime.\cite{mu1992} 
As the pumping rate crosses the critical point $W^*$ the stationary correlator $g^{(2)}(0)$ exhibits an abrupt growth, while the correlation lifetime $\tau^{(2)}$ demonstrates non-monotonous behavior with a sharp peak. It rises as $\tau^{(2)}\propto 1/|W-W^*|$ near the critical point $W=W^*$. The peak height  is on the order of $g/\Gc^2$, and much larger than the value of the correlation time  in all other regimes.

At large pumping $W \gg W^*$ the strong coupling regime is destroyed: the  emission statistics is thermal [$g^{(2)}(0)=2$] and the decay time of the correlations equals to the empty cavity mode lifetime $1/\Gc$.

Now we proceed to more detailed analysis of the regimes (A)--(C).
\subsection{Low pumping, $W\ll \Gc$}\label{sec:low}
The density matrix equations can be conveniently analyzed using the basis of the 
 eigenstates of the Hamiltonian~(\ref{eq:ham}), which are well defined in strong coupling regime.
The eigenstates read\cite{jaynes1963}
\begin{equation}\label{eq:basis}
\ket{0} = \ket{0,G} \,, \:\: \ket{m,\pm}=\frac{\ket{m,G}\pm\ket{m-1,X}}{\sqrt{2}}\:\: \:(m\geq1)\:,
\end{equation}
where $\ket{m,G}$ and $\ket{m,X}$ stand for the states with $m$ photons and no excitons or one exciton, respectively. The energy spectrum forms the  Jaynes-Cummings ladder
\begin{equation}
 E_0=0,\quad E_{m,\pm}=m\hbar\omega_0 \pm \sqrt{m}\hbar g\:. \label{eq:ladder}
\end{equation}
Each rung of the ladder contains two states split by the Rabi frequency   $2\sqrt{m}g/\hbar$, increasing with the rung number $m$.

In the limit of vanishing pumping $W \ll\Gc$ it is sufficient to take into account only the rungs with $m\le 2$ particles, which yields the following  correlation function,
\begin{equation}\label{eq:g2t_strong_linear}
g^{(2)}(t)=1-\frac{3\Gc-\Gx-(\Gc+\Gx)\cos 2gt}{2(3\Gc+\Gx)} \, \e^{-(\Gc+\Gx)t/2}\:.
\end{equation}
Eq.~\eqref{eq:g2t_strong_linear} shows the oscillations of the photon-photon correlator. This is the direct manifestation of the strong coupling regime. Due to the photon-exciton interaction $g$ the photon is fully converted into exciton and vice versa every period $\pi/(2g)$ (Rabi oscillations). This results in the contribution to the  correlator, oscillating with the Rabi frequency of the first rung $(E_{1,+}-E_{1,-})/\hbar=2g$. The decay rate of the Rabi oscillations
 is the average of the photon and exciton decay rates $(\Gc+\Gx)/2$. This is also a manifestation of strong coupling and shows formation of the excitonic polaritons.
For realistic cavities $\Gc\gg\Gx$,\cite{Khitrova2006}  which means that $g^{(2)}\approx 2/3<1$  (antibunching).\cite{Valle2011}
The black dotted curve in the Fig.~\ref{fig:strong}(a) is plotted after Eq.~\eqref{eq:g2t_strong_linear} and well reproduces the numerical results for low pumping rate (black solid curve).

\subsection{Moderate pumping, $\Gc \lesssim W \ll g^2/\Gc$}\label{sec:moderate}
In contrast to the low pumping case (Sec.~\ref{sec:low}), for  moderate pumping  it is necessary to  take into account all of the rungs. First, one needs to determine the  stationary density matrix $\rho^{(0)}$.  As shown in Appendix~\ref{ApS} this matrix $\rho^{(0)}$ is diagonal in the basis of states~(\ref{eq:basis}) and has the elements $\rho^{(0)}_{0;0}=f^{(0)}_0$, $\rho^{(0)}_{m,\pm;m,\pm} = f^{(0)}_m$, where the distribution function $f^{(0)}_m$ reads
\begin{equation}\label{eq:distr}
f^{(0)}_m = \frac{\sqrt{\pi}}{2E(w)+1} \: \frac{ w^m}{\Gamma(m+1/2)} \:.
\end{equation}
Here, $w=W/2\Gc$ is the dimensionless pumping, $E(w)=  \e^w \sqrt{\pi w} \, {\rm erf }\sqrt{w}$, $\Gamma(x)$ and ${\rm erf}(x)$ are the gamma and error functions, respectively.
For simplicity in this section the exciton decay rate $\Gx$ is neglected, since for typical microcavities $\Gx\ll\Gc$.\cite{Khitrova2006,Forchel2010} Equation~\eqref{eq:distr} generalizes the analytical result for the distribution function obtained in Ref.~\onlinecite{Poddubny2010prb}.

The distribution Eq.~\eqref{eq:distr} provides the following expressions for the particle numbers and the stationary correlator $g^{(2)}(0)$:
\begin{align}\label{eq:Ng2_strong}
&\Nx= \frac{E}{2E+1} \:,\qquad \Nc = 2w \,\frac{E+1}{2E+1} \:, \\
&g^{(2)}(0) = \frac{(2E+1)[4w^2(E+1)+E-2w]}{8w^2(E+1)^2} \nonumber \:.
\end{align}
\begin{figure}[ht]
\includegraphics[width=.99\columnwidth]{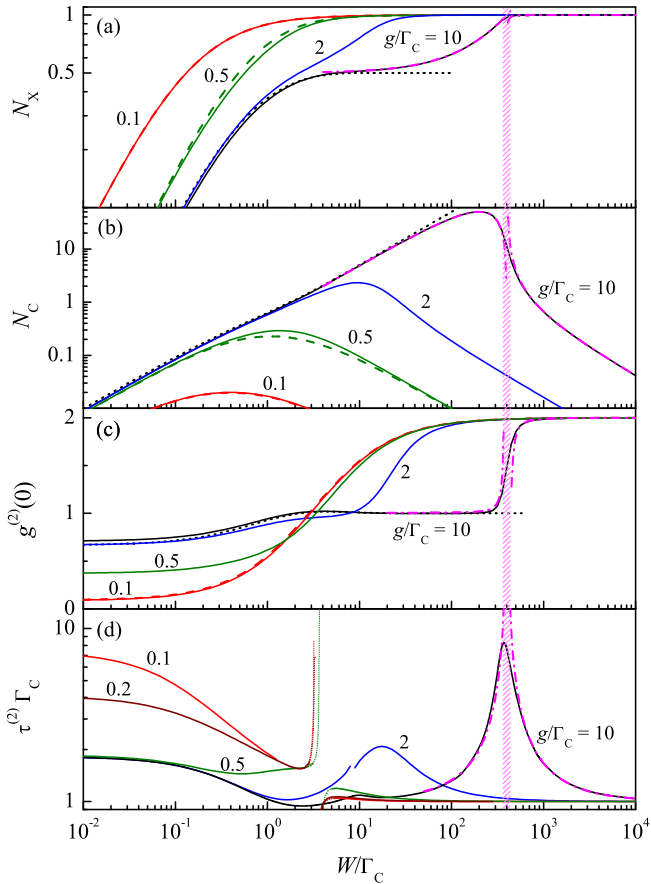}
\caption{(Color online) Dependence on the pumping rate $W$ of (a) exciton number $\Nx$, (b) photon number $\Nc$, (c) correlation function $g^{(2)}(0)$, and (d) correlation decay time $\tau^{(2)}$, plotted for various coupling strengths $g$ and $\Gx=0.1\Gc$. Solid lines are obtained from numerical calculation. Dashed lines in panels (a)-(c) present weak coupling regime and are plotted after Eqs.~(\ref{Nxc_weak}) and (\ref{g20_weak}), while dotted lines correspond to the strong coupling limit, Eqs.~(\ref{eq:Ng2_strong}). Dash-dotted lines  in panels (a)-(d) show the analytic results obtained at high pumping regime Eq.~\eqref{QNcNxg20} and~\eqref{Qtaug2} and valid outside the patterned region $|W-W^*|\sim 4g = 40 \Gc$ for the parameters chosen. 
Thin dotted lines in panel (d) correspond to the range of pumping, where the function $g^{(2)}(t)$ is close to $1$ and the definition of $\tau^{(2)}$ according to Eq.~\eqref{eq:decay_time} fails.
}
\label{fig:static}
\end{figure}
Solid curves in the Figs.~\ref{fig:static}(a)--\ref{fig:static}(c) show the dependence of $\Nx$, $\Nc$ and  $g^{(2)}(0)$ on the pumping rate. The curves are presented at the different values of exciton-photon coupling strength $g$. 
Results in Figs.~\ref{fig:static}(a)--\ref{fig:static}(c) agree with those obtained numerically in Ref.~\onlinecite{laussy2009}. 
Here we focus on the  strong coupling regime (black curves),  the weak coupling case will be analyzed in Sec.~\ref{sec:weak}.
The dot population $\Nx$ [see Fig.~\ref{fig:static}(a)] monotonously increases at low pumping  as $\Nx=W/\Gc$ and reaches the plateau $\Nx=1/2$ at $W \sim \Gc$.  
The plateau $\Nx=1/2$ reflects the half-exciton half-photon nature of the polariton eigenstates Eq.~\eqref{eq:basis} with $m\ge 1$.
Photon number $N_{\rm C}$ [see Fig.~\ref{fig:static}(b)] linearly grows for low and moderate pumping as $\Nc=W/\Gc$ and  $\Nc=W/(2\Gc)$, respectively. 
The second-order correlator $g^{(2)}(0)$ [see Fig.~\ref{fig:static}(c)] grows with pumping and reaches the plateau $g^{(2)}(0)=1$ at $W \sim \Gc$ (lasing regime\cite{mu1992,Poddubny2010prb}).
In the moderate pumping regime all the curves $\Nx$, $\Nc$ and $g^{(2)}(0)$ are well described by Eq.~(\ref{eq:Ng2_strong}), see black dotted curves. 

Now we proceed to the discussion of the  dynamics of $g^{(2)}(t)$.
 Two independent contributions can be singled out in the time dependence, see Fig.~\ref{fig:strong}(a): (i) the oscillatory contribution and (ii) the
monotonously decaying contribution.
As shown in Appendix~\ref{ApS},  the oscillatory term
demonstrates the superposition of the Rabi beatings between the split states inside different rungs of Jaynes-Cummings ladder with the frequencies $2\sqrt{m}g$. The weight of the term corresponding to the rung  $m$ is determined by the distribution function $f^{(0)}_{m+1}$. The damping of the oscillations is due to the stimulated photon decay and exciton pumping and equals to $\Gc(m-1/2)+W/2$. 
 Non-oscillating term decays in time towards unity on the time scale of $\Gc$.
Calculation shows that this decay can be approximated as exponential one with the rate $\Gamma_1$,
\begin{equation}\label{eq:BigEquation}
\Gamma_1 = \Gc \,\left[1-\frac{2 w-1}{4w(E+1)}-\frac{(2 w-1)^2}{4w(2E+1)}\right]^{-1}\:.
\end{equation}
For small pumping $W \ll \Gc$ Eq.~\eqref{eq:BigEquation} reduces to $\Gamma_1 = \Gc/2$, whereas for moderate pumping $W \gg \Gc$ one gets $\Gamma_1 = \Gc$. Resulting expression for the correlator $g^{(2)}(t)$ assumes the form
\begin{align}\label{eq:g2t_strong}
g^{(2)}(t) = &1-\frac{1}{2(E+1)}\,\e^{-\Gamma_1 t}\\
 + &\frac{1}{2 \Nc^2} \sum\limits_{m=1}^{\infty} f^{(0)}_{m+1} \cos(2\sqrt{m}gt) \, \e^{-[\Gc(m-1/2)+W/2]t} \:. \nonumber
\end{align}
For low pumping the sum in Eq.~(\ref{eq:g2t_strong}) is determined by the first term with $m=1$ and the result agrees with Eq.~(\ref{eq:g2t_strong_linear}) assuming $\Gx=0$.

Analytical results plotted after Eq.~(\ref{eq:g2t_strong}) are shown in the Fig.~\ref{fig:strong}(a) by dashed curves. The difference from the exact calculation at small pumping is due to the neglected exciton decay rate $\Gx$.  
Eq.~\eqref{eq:g2t_strong} well reproduces the main features of the numerically calculated dependence: for larger pumping the amplitude  of the oscillations significantly decreases and they decay faster.
This is a characteristic feature of the two-level system, distinct from the  bosonic system where the lifetime of fluctuations increases with pumping.\cite{Haken1964,Glazov2013}

\subsection{High pumping, $W \sim 4g^2/\Gc$}\label{sec:high}

When the pumping rate is increased up to $W \sim g^2/\Gc$ the exciton level broadening 
 caused by pumping
becomes comparable to the rung splitting. This results in 
 saturation of exciton number at unity [see Fig.~\ref{fig:static}(a)] and drastic decrease of the photon number [see Fig.~\ref{fig:static}(b)].

 The detailed description of stationary density matrix and correlator dynamics equations is given in Appendix~\ref{ApQ}. It is shown that the emission statistics changes qualitatively  when the pumping rate crosses the critical value $W^* = 4g^2/\Gc$. 
For lower than critical pumping the distribution function is Gaussian, while for larger pumping it becomes  thermal. The transition occurs in the vicinity of critical point  $|W-W^*| \lesssim 4g$. Below we present the analytical expressions for emission characteristics valid outside this narrow region.

The static characteristics for $|W-W^*| \gtrsim 4g$ are given by
\begin{align}\label{QNcNxg20}
\Nx&= \begin{cases}
      \dfrac{1+W/W^*}{2}, & (W<W^*) \\
      1-\dfrac{\Gc}{W(W/W^*-1)}, &  (W>W^*) \:,
    \end{cases} \\
\Nc&= \begin{cases}
      \dfrac{W}{2\Gc}\, (1-W/W^*), & (W<W^*) \\
      \dfrac{1}{W/W^*-1}, & (W>W^*)\:,
    \end{cases} \nonumber\\
g^{(2)}(0)&= \begin{cases}
      1+\dfrac{2\Gc}{W^*(1-W/W^*)^2}, & (W<W^*) \\
      2-\dfrac{4\Gc}{W(1-W^*/W)^2}, & (W>W^*)\:.
    \end{cases} \nonumber
\end{align}
Dash-dotted lines in Figs.~\ref{fig:static}(a)--\ref{fig:static}(c) present the dependence of $\Nx$, $\Nc$ and  $g^{(2)}(0)$ on the pumping rate near the critical point $W^*$ plotted after Eq.~\eqref{QNcNxg20}. One can see the perfect agreement of the analytical results with the numerical calculation (black solid curve) outside the narrow patterned transition region.

The time dependence of the correlator $g^{(2)}(t)$ is mono-exponential,
\begin{equation}\label{Qg2t}
g^{(2)}(t) = 1+ [g^{(2)}(0)-1] \,\e^{-t/\tau^{(2)}}  \:,
\end{equation}
where the correlation lifetime is given by
\begin{equation}\label{Qtaug2}
\tau^{(2)} = \frac{1}{\Gc} \times \,\begin{cases}
      \dfrac{1}{1-W/W^*}, & (W<W^*) \:, \\
      \dfrac{1}{1-W^*/W}, & (W>W^*) \:.
    \end{cases} 
\end{equation}
Eq.~\eqref{Qtaug2} shows that the correlation lifetime drastically grows near the critical point $W=W^*$. Its maximum value can be estimated from Eq.~\eqref{Qtaug2} by substituting $|W-W^*| = 4g$, which gives the value $\tau^{(2),\rm max} = g/\Gc^2$ that is larger than the correlation lifetime in all other regimes by the factor $g/\Gc$. The dependence of the lifetime  $\tau^{(2)}$ on the pumping rate near the critical point $W^*$ is shown in the Fig.~\ref{fig:static}(d) by the dash-dotted line.

The origin of the peak in the correlation lifetime at $W=W^*$ can be qualitatively understood as follows. At $W>W^*$, the strong coupling regime is already destroyed due to the self-quenching. However, if $W-W^* \ll  W^*$, the number of photons in the cavity is still large, see Eq.~\eqref{QNcNxg20}. Hence, this system can be viewed as a conventional weak-coupled laser, where the fluctuations lifetime is increased due to the bosonic stimulation factor, $\tau^{(2)}=(\Nc+1)/\Gc$.\cite{Haken1964,Glazov2013} For single dot in the cavity such decay time enhancement  can be realized only in strong coupling, because in weak coupling case the number of photons remains small at any pumping, see Fig.~\ref{fig:static}(b) and Sec.~\ref{sec:weak} below.
\section{Weak coupling regime}\label{sec:weak}
\begin{figure}
\includegraphics[width=.99\columnwidth]{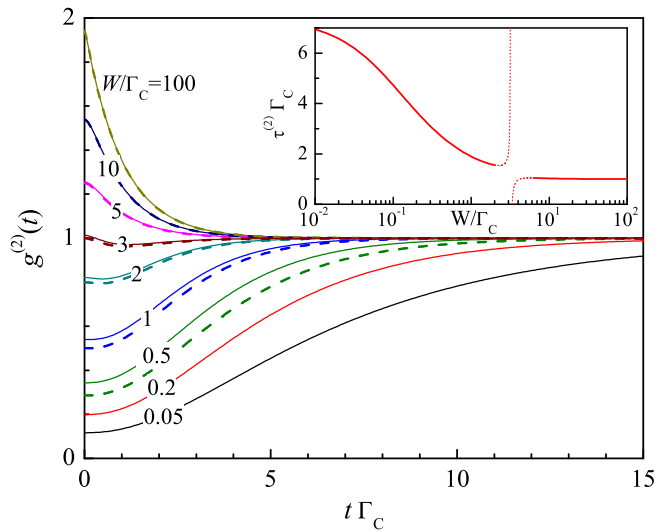}
\caption{(Color online) Time dependence of the correlation function $g^{(2)}(t)$ in the weak coupling regime, $g/\Gc=0.1$. Curves are plotted for $\Gx=0.1\Gc$ and various pumping rates $W/\Gc$. Solid lines are obtained from the numerical calculation.
Dashed lines present moderate pumping case and are plotted after Eq.~(\ref{g2t_weak}).
The inset shows the pumping dependence of the correlation lifetime $\tau^{(2)}$.
Dashed region corresponds to the unphysical divergence in $\tau^{(2)}$ near $W/\Gc = 3$ when
$g^{(2)}(t)\approx 1$ and the lifetime Eq.~\eqref{eq:decay_time} is of no sense.}
\label{weak}
\end{figure}
In this section we analyze the structures where the strong coupling condition $g \gg \Gc,\Gx$ is violated. Figures~\ref{fig:static}(a)--\ref{fig:static}(d) show the dependence of the particle numbers, stationary two-photon correlator and correlation lifetime on the coupling strength. 
Decrease of the coupling strength parameter $g/\Gc$ suppresses the maximal number of photons and  peak value of $\tau^{(2)}$, and also shifts the self-quenching transition  to the lower values of pumping (blue curves in Fig.~\ref{fig:static}). As soon as the coupling strength $g$ becomes smaller than  $\Gc$, the  regime of weak coupling between the photon and the exciton is realized. At weak coupling the number of photons is much less than one  at any pumping as can be clearly seen from the red curve ($g/\Gc=0.1$) in Fig.~\ref{fig:static}(b). 

Smallness of the photon number allows one to consider only the lowest levels of the system when deriving analytical results. Hence, we take into account the states with no more than one photon when  calculating photon number and dot occupation, and the states with up to two photons for photon-photon correlator.
Analytical expressions for the particle numbers in the weak coupling regime read\cite{Laussy2009B}
\begin{align}
\Nx = \frac{W [4g^2+\Gc (W+\Gc+\Gx)]}{(W+\Gc +\Gx)[4g^2+\Gc (W+\Gx)]} \label{Nxc_weak}\:,\\ 
\Nc = \frac{4g^2 W}{(W+\Gc +\Gx)[4g^2+\Gc (W+\Gx)]} \label{Nc_weak}\:. \nonumber
\end{align}
At low pumping both $\Nx$ and $\Nc$ grow linearly with pumping. At large pumping the dot is completely populated, $\Nx=1$, while the cavity is empty ($\Nc\to 0$) due to the self-quenching  effect, see Figs.~\ref{fig:static}(a) and~\ref{fig:static}(b).

Analytical expression for  $g^{(2)}(0)$ in the ``bad'' cavity regime ($g \ll \Gc$) is
\begin{equation}\label{g20_weak}
g^{(2)}(0)=2\,\frac{W+\Gx+ 4g^2/\Gc}{W+\Gx+3\Gc} \:.
\end{equation}
In the limit of vanishing pumping and $\Gx\ll\Gc,g^2/\Gc$ the value of $g^{(2)}(0)$ is smaller than the strong coupling limit $2/3$.
With decrease of the coupling strength $g$ antibunching  becomes stronger due to  smaller admixture of photons to the exciton state.

With increase of the pumping rate the initial value $g^{(2)} (0)$ grows from zero (antibunching) to $2$ (thermal regime), see Fig.~\ref{fig:static}(c). The lasing regime with the plateau at $g^{(2)} (0)=1$ is destroyed in weak coupling case.
Shown in the Figs.~\ref{fig:static}(a)--\ref{fig:static}(c) by the dashed lines is the analytical dependence plotted after Eqs.~(\ref{Nxc_weak}) and~(\ref{g20_weak}). One can see the perfect agreement with the numerical calculation shown by the solid lines.

The time dependence $g^{(2)}(t)$ is calculated according to the procedure defined by Eq.~(\ref{eq:g2_eq}). 
For simplicity we neglect the exciton damping $\Gx=0$ and consider only the two limiting cases of  pumping low and high as compared to the spontaneous decay rate of the exciton $4g^2/\Gc$. At low pumping we obtain
\begin{equation}
g^{(2)}(t) = 1 - \e^{-4g^2 t/\Gc}\:.
\end{equation}
The system demonstrates antibunching similar to the case of the quantum dot without the cavity. The only effect of the cavity is the enhancement of the exciton decay rate due to the Purcell effect.
In the opposite case of high pumping, $W \gg g^2/\Gc$, one gets
\begin{align}\label{g2t_weak}
g^{(2)}&(t) = 1 - \frac{\Gc^2}{(W-\Gc)^2} \left[ \frac{2W(W-5\Gc )}{\Gc (W+3\Gc)}\e^{-(W+\Gc) t/2} \right. \nonumber\\
&+ \left. \frac{W(2\Gc^2 +3W\Gc-W^2)}{\Gc^2 (W+3\Gc )} \e^{-\Gc t} + \e^{-W t}\right] \:.
\end{align}
We note that Eq.~(\ref{g2t_weak}) is finite at $W=\Gc$ and reduces to $1-\e^{-\Gc t}(\Gc^2 t^2+2\Gc t+2)/4$ for this particular value of pumping. In the Fig.~(\ref{weak}) the dependence defined by Eq.~(\ref{g2t_weak}) is plotted by dashed curves. For the high pumping $W \gg \Gc$  Eq.~(\ref{g2t_weak}) reduces to $g^{(2)}(t) = 1+\e^{-\Gc t}$. This corresponds to low number of cavity photons and thermal statistics.
The decay time $\tau^{(2)}$ decreases  from  $\Gc/(4g^2)$ to $1/\Gc$ with pumping growth, see Fig.~\ref{fig:static}(d) and inset in Fig.~\ref{weak}. 
Although the behavior of the decay time in weak coupling is generally monotonous, there is a region of pumping values, where $g^{(2)}(t)$ is close to unity, see the wine-colored curve for $W/\Gc=3$.
In this case 
the correlation lifetime $\tau^{(2)}$ defined according to Eq.~\eqref{eq:decay_time} is of no sense.  Thin dotted lines in Fig.~\ref{fig:static}(d) and in the inset of   Fig.~\ref{weak} correspond to this region.

\section{Summary}\label{sec:summary}
Theory of time-resolved second-order correlations of photons, emitted from the incoherently stationary pumped microcavity with single quantum dot has been developed. Explicit analytical expressions for the photon number, exciton number, and the photon-photon correlator $g^{(2)}(t)$ have been obtained.
In the strong coupling regime the correlation function $g^{(2)}(t)$ demonstrates decaying in time Rabi oscillations. Both the frequency and the decay rate of these oscillations increase with growth of the pumping rate. 
At larger pumping the dynamics of the correlations is monoexponential. The decay time nonmonotonously depends on the pumping and has a sharp peak at the critical pumping value corresponding to the self-quenching transition between lasing regime [where $g^{(2)}(0)=1$] and the thermal regime [$g^{(2)}(0)=2$]. The peak value strongly exceeds the lifetime of the empty cavity mode. Such nonmonotonous behavior of the correlation lifetime is a characteristic feature of the cavity with single dot in the strong coupling regime.

 In the weak coupling regime the correlation function almost monotonously changes from the  initial value at $t=0$ to unity at large delays. 
 The value of zero delay correlator $g^{(2)}(0)$ in weak coupling regime is smaller than unity at low pumping (photon antibunching) and tends to 2 at large pumping (thermal bunching). The increase of pumping shortens the  decay time of the photon-photon correlations.

\paragraph*{Acknowledgments.} The authors acknowledge fruitful discussions with
M.M. Glazov. This work was supported by the RFBR, 
RF President Grants MD-2062.2012.2 and NSh-5442.2012.2, EU projects SPANGL4Q and
POLAPHEN, and the ``Dynasty'' Foundation.

\appendix
\section{Dynamic equations in the strong coupling regime}\label{ApS}
In this Appendix we present the details of the derivation of the analytical answers 
Eqs.~\eqref{eq:distr}--\eqref{eq:g2t_strong} for the stationary density matrix and for  the time-dependent two-photon correlator. We focus on the strong coupling regime and moderate pumping $W\ll 4g^2/\Gc$.

The key simplification in the strong coupling regime is the smallness of the non-diagonal components of the stationary density matrix. This is valid because the energy width of the polariton eigenstates Eq.~\eqref{eq:basis} is on the order of $ \max(\Gc,W)$ and much less than the splitting between these states $2\sqrt{m}g$,
where  $m$ is the relevant rung number of the Jaynes-Cummings ladder.  Estimating the typical values of $m$ as $W/\Gc$ we obtain  the small parameter $\max (\Gc, \sqrt{W \Gc}) / g$ for the non-diagonal  density matrix elements. Note, that for sufficiently high pumping $W \sim g^2/\Gc$ this parameter is no longer small. The non-diagonal density matrix elements in this case are given by Eq.~\eqref{Qx_m} and their effect is discussed in detail in Appendix~\ref{ApQ}.
 
Thus, in the regime of moderate pumping one can consider the dynamics of the diagonal and non-diagonal density matrix elements separately. The kinetic equation for the diagonal  elements $f_m=\rho_{m,+;m,+}=\rho_{m,-;m,-}$, $f_0=\rho_{0;0}$ is obtained from Liouvillian Eq.~\eqref{liou} and reads
\begin{align}\label{eq:Adiag_eq}
\frac{\rmd f_m}{\rmd t} =& -\frac{W}{2}(f_m-f_{m-1})-\frac{\Gx}{2}(f_m - f_{m+1}) \nonumber\\
 &- \Gc[(m-1/2)f_m-(m+1/2)f_{m+1}]\:,  \nonumber\\
\frac{\rmd f_0}{\rmd t} =& -W f_0 + (\Gx + \Gc) f_1\:.
\end{align}
Stationary solution of this equation can be found using the fact that the probability flow between the two adjacent rungs $m$ and $m+1$ should be zero, i.e.
\begin{equation}
-\frac{W}{2}f^{(0)}_m+\frac{\Gx}{2}f^{(0)}_{m+1}+\Gc(m+1/2)f^{(0)}_{m+1} = 0 \:.
\end{equation}
This yields the following probability distribution,
\begin{equation}\label{eq:AdistrGx}
f^{(0)}_m \propto \frac{(W/2\Gc)^m}{\Gamma(m+1/2+\Gx/2\Gc)} \:,
\end{equation}
that should be normalized according to
\begin{equation}\label{eq:AdistrNorm}
f^{(0)}_0 + 2\sum_{m=1}^{\infty} f^{(0)}_m = 1 \:.
\end{equation}
Hereinafter  we neglect the exciton damping for simplicity, $\Gx=0$. In this case  Eqs.~(\ref{eq:AdistrGx})~and~(\ref{eq:AdistrNorm}) lead to Eq.~(\ref{eq:distr}).
Stationary photon ($\Nc$) and exciton ($\Nx$)  numbers and 
the two-photon correlator
$g^{(2)}(0)$ are  readily found from the distribution  $f_m^{(0)}$,
\begin{align}
&\Nx= \sum\limits_{m=1}^{\infty} f^{(0)}_m \:,\qquad \Nc = \sum\limits_{m=1}^{\infty} f^{(0)}_m (2m-1)\:, \\
&g^{(2)}(0) = \frac{2}{\Nc^2} \sum\limits_{m=2}^{\infty} f^{(0)}_m (m-1)^2\:.
\end{align}
which yields Eqs.~(\ref{eq:Ng2_strong}).

The time dynamics of $g^{(2)}(t)$ is governed by Eq.~(\ref{eq:g2_eq}).
One needs to determine the time dependence of the operator $\chi (t)\equiv {\rm e}^{\mathcal L t}[c \rho_0c^\dag]$. This operator can be separated into diagonal and nondiagonal parts in the basis of polarion eigenstates: $\chi_{0;0}=\chi_0^{\rm (d)}$, $\chi_{m,\pm;m,\pm}=\chi_m^{\rm (d)}$ (diagonal part), and $\chi_{m,+;m,-}=\chi_{m,-;m,+}^*=\chi_m^{\rm (nd)}$ (nondiagonal part). These two parts evolve independently in the  regime of strong coupling and moderate pumping. Correlation function  Eq.~(\ref{eq:g2_def}) can be expressed through the matrix elements as
\begin{equation}\label{Ag_chi}
g^{(2)}(t) = \frac{1}{\Nc^2} \sum\limits_{m=1}^{\infty} \left[ (2m-1)\chi^{\rm (d)}_m(t) + \Re \chi^{\rm (nd)}_m(t) \right] \:.
\end{equation}
Initial conditions at $t=0$ read
\begin{align}
\chi^{\rm (d)}_0(0) = f_1^{(0)}\:,\quad  &\chi^{\rm (d)}_m(0)=(m+1/2)f^{(0)}_{m+1} \:,\\
&\chi^{\rm (nd)}_m(0)=f^{(0)}_{m+1} /2 \qquad (m\geq 1)\:.  \nonumber
\end{align} 
Below we  consider the time dynamics of the diagonal matrix elements first, and then analyze the non-diagonal ones.

Diagonal elements satisfy  Eqs.~(\ref{eq:Adiag_eq}) where $f_m$ should be replaced with $\chi^{\rm (d)}_m$. To determine their time dependence  we analyze the eigenvectors and eigenvalues of this linear system. The largest eigenvalue is zero and corresponds to the stationary distribution function $f_{m}^{(0)}$. All other eigenvalues are negative and describe the solutions decaying with time. 
Our goal is to provide an estimation for the non-zero eigenvalue with the smallest absolute value 
$\Gamma_1$. Comparison with the numerical calculation demonstrates that this single eigenvalue satisfactory describes the  dynamics of the diagonal matrix elements $\chi^{\rm (d)}(t)$.

The Lindblad-type matrix, corresponding to the right-hand side of Eqs.~(\ref{eq:Adiag_eq}), is not Hermitian. The problem can be still reduced to a Hermitian one by the procedure adopted for kinetic and Fokker-Planck equations.\cite{Risken_1989} In our case this procedure is formally equivalent to the  definition of the scalar product of two distributions $u_m$ and $v_m$ as
\begin{equation}
 (u,v)\equiv \frac{u_0 v_0}{f^{(0)}_0}+2\sum\limits_{m=1}^\infty \frac{u_m v_m}{f_m^{(0)}}\:. \label{eq:scalar}
\end{equation}
In terms of the scalar product~(\ref{eq:scalar}) the operator corresponding to right-hand side of Eqs.~\eqref{eq:Adiag_eq},  denoted in what follows by dot, turns out to be self-adjoint, i.e. $(u,\dot{v})=(\dot{u},v)$ for any distributions $u_m$ and $v_m$.  This allows one to analyze kinetic Eqs.~\eqref{eq:Adiag_eq} by the standard approaches of quantum mechanics. For instance, the normalization condition Eq.~\eqref{eq:AdistrNorm} assumes the compact form $(f^{(0)},f^{(0)})=1$.  

It is convenient to single out from the considered distribution $\chi^{\rm (d)}_m$ the stationary contribution $(f^{(0)},\chi^{\rm (d)})\,f^{(0)}_m$, that corresponds to zero eigenvalue. This yields
\begin{equation}
 \chi^{\rm (d)}_m(t)=\Nc f^{(0)}_m+\delta \chi^{\rm (d)}_m(t)\:.\label{eq:ansatz}
\end{equation}
Initial conditions for the new variables $\delta \chi^{\rm (d)}_m(t)$ read
\begin{equation}
 \delta\chi^{\rm (d)}_0 (0) = \frac{2E w f^{(0)}_0}{2E+1} \:,\quad \delta\chi^{\rm (d)}_m (0) = -\frac{w f^{(0)}_m}{2E+1} \:.\label{eq:chi_d_0}
\end{equation}
Since the projection of  $\delta \chi^{\rm (d)}$  onto the stationary solution $(\delta\chi^{\rm (d)},f^{(0)})$ is zero, $\delta\chi^{(d)}(t)$ vanishes at large times, which  allows to verify  that $g^{(2)}(t)\to 1$ at $t\to \infty$.
Approximating the time decay as
$\delta\chi^{(d)}(t)=\delta\chi^{(d)}(0)\,\e^{-\Gamma_1t}$ and substituting
Eqs.~\eqref{eq:ansatz},\eqref{eq:chi_d_0} into Eq.~\eqref{Ag_chi} we obtain the first two terms in the right-hand side of Eq.~\eqref{eq:g2t_strong}. Here, $\Gamma_1$  is the  desired eigenvalue governing the time decay of $\delta\chi^{\rm (d)}$. It can be estimated using the following variational ansatz for the corresponding eigenvector
\begin{equation}\label{eq:f1}
 f_{m}^{(1)}=f_{m}^{(0)}(1+\alpha m) \qquad (m \geq 0) \:.
\end{equation}
The constant $\alpha$ is found by imposing the orthogonality condition $(f^{(1)},f^{(0)})=0$.
Once $\alpha$ is found, the value of $\Gamma_1$ is given by
\begin{equation}
 \Gamma_1=\frac{(f^{(1)},\dot f^{(1)})}{(f^{(1)},f^{(1)})}\:.\label{eq:Gamma1}
\end{equation}
The result of calculations reduces to Eq.~\eqref{eq:BigEquation}.

Now we turn to the evolution of the nondiagonal matrix elements $\chi_m^{(\rm nd)}$.
In the strong coupling regime this procedure is quite simple.
One can assume that each matrix element $\chi^{\rm (nd)}_m(t)$ oscillates in time with its own frequency $(E_{m,+}-E_{m,-})/\hbar = 2\sqrt{m}g$ and decays exponentially. The intermixing of different rungs
can be neglected  provided that the frequency difference between the adjacent rungs $E_{m+1,+}-E_{m,+}\sim g(\sqrt{m+1}-\sqrt{m})$ is smaller than the coupling term, which is on the order of the pumping strength $W$.
This is realized for $W \ll g^{2/3}\Gc^{1/3}$, i.e. below the transition to the  lasing regime.~\cite{Poddubny2010prb}
Note that in the polariton lasing regime $g^{(2)}(0)=1$ [see Eq.~(\ref{eq:Ng2_strong})], and the photon correlator dynamics is trivial, $g^{(2)}(t) \equiv 1$. Therefore this case does not need any special consideration.
Hence, the desired time dependence of $\chi^{\rm (nd)}_m$ is described as
\begin{equation}\label{eq:And_t}
\chi^{\rm (nd)}_m(t) = \chi^{\rm (nd)}_m(0) \: \e^{-2\i\sqrt{m}gt} \, \e^{-[\Gc(m-1/2)+W/2]t}\:.
\end{equation}
Substituting the non-diagonal component dynamics defined by Eq.~(\ref{eq:And_t}) into Eq.~(\ref{Ag_chi}) we recover the last term in Eq.~(\ref{eq:g2t_strong}).

\section{Dynamic equations in the self-quenching regime}\label{ApQ}
In this Appendix we present the details of the  $g^{(2)}(t)$ correlator dynamics for the system in strong coupling regime and under high pumping. We focus on the transition from lasing to self-quenching regime ($W \sim 4g^2/\Gc$) when the correlation lifetime turns out to be extremely long.

Despite the strong coupling,  sufficiently high pumping leads to the intermixing of the Hamiltonian Eq.~\eqref{eq:ham} eigenstates. Therefore the density matrix is no more diagonal
in the polariton basis. Inside the $m$-th rung of Jaynes-Cummings ladder it can be written as
\begin{equation}\label{Qdenm}
\rho_m =   \left( 
\begin{array}{cc} 
f_m & x_m \\ 
x_m^* & f_m 
\end{array} 
\right) \:
\end{equation}
in the basis of eigenstates Eq.~\eqref{eq:basis}\:.
The Liouvillian~\eqref{liou} does not mix the intra-rung and inter-rung density matrix components. Thus, the stationary density matrix, as well as the operator $\chi (t)$ that describes  $g^{(2)}(t)$ dynamics, do not contain inter-rung components.

In the considered regime of high pumping the density matrix equations for the rung $m$ read
\begin{align}\label{Qdyn1}
\frac{\rmd f_m}{dt} &= \frac{W}{2}[f_{m-1} - f_m + \Re(x_{m-1} - x_m)]    \nonumber\\
 &+ \Gc [(m+1/2)f_{m+1} - (m-1/2)f_{m}]  \:,\nonumber\\
\frac{\rmd x_m}{\rmd t} &= -2\i \sqrt{m} g x_m - W(f_m +3x_m/4 + x_m^*/4)  \:.
\end{align}
The diagonal components $f_m$  change with the rate  of the order of $\Gc$, as will be proven later. The non-diagonal component $x_m$ relaxes to its quasi-stationary value with much larger rate on the order of $W$. Hence, we assume that $x_m$ adiabatically follows diagonal components $f_m$, i.e.
\begin{equation}\label{Qx_m}
 x_m = \frac{2\i \sqrt{m} gW - W^2/2}{4mg^2 + W^2/2} \, f_m \:.
 \end{equation} 
After substitution of the expression Eq.~\eqref{Qx_m} for the non-diagonal components  into Eq.~\eqref{Qdyn1} we obtain
\begin{align}
\frac{\rmd f_m}{\rmd t} &= \frac{W}{2}(\xi_{m-1} f_{m-1} - \xi_{m} f_m) \nonumber\\
 &+ \Gc [(m+1/2)f_{m+1} - (m-1/2)f_{m}]\:,\label{eq:fq}
\end{align}
where $\xi _m =m/(m+q)$ and $q = W^2/(8g^2)$. In moderate pumping case 
$\xi_m$ is close to unity and
Eq.~\eqref{eq:fq} reduces to Eq.~\eqref{eq:Adiag_eq}.
 
Since we consider high rungs we can replace discrete rung number $m$ with continuous variable, which yields 
\begin{align}\label{QdynGenral}
&\frac{\partial f(m,t)}{\partial t} + \frac{\partial j(m,t)}{\partial m}  = 0 \:,\\
&j(m,t) =  \Gc m \left[\frac{w-q-m}{m+q} \,f(m,t) -  \frac{\partial f(m,t)}{\partial m} \right] \nonumber\:,
\end{align}
where $j(m,t)$ is the probability current. 
The stationary solution found from the condition $j(m)=0$ reads
\begin{equation}\label{Qf0}
f^{(0)} (m) \propto (m+q)^w \e^{-m}\:.
\end{equation}
With the growth of the pumping the maximum of the stationary distribution function Eq.~\eqref{Qf0} behaves as $w-q$. It grows linearly at low pumping, reaches maximum at $W=2g^2/\Gc$, then decreases, reaches zero at critical pumping value $W^*=4g^2/\Gc$ when $w=q=2g^2/\Gc^2$, and remains zero for higher pumping. 

First we consider the subcritical pumping, $W<W^*$. The stationary distribution function Eq.~\eqref{Qf0} in this case can be approximated as Gaussian,
\begin{equation}\label{Qf0left}
f^{(0)} (m) = \frac{1}{\sqrt{8\pi w}} \, {\rm exp}\left[-\frac{(m-w+q)^2}{2w} \right] \:.
\end{equation}
We use the distribution function Eq.~\eqref{Qf0left} for $w-q \gg \sqrt{2w}$ to calculate $\Nx$, $\Nc$, and $g^{(2)}(0)$ according to Eqs.~\eqref{eq:NcNx_def} and~\eqref{eq:g20_def} and taking into account both diagonal and non-diagonal components of the stationary density matrix Eq.~\eqref{Qdenm}. This yields the results  presented in the upper parts of Eqs.~\eqref{QNcNxg20}.
Time-dependent Eq.~\eqref{QdynGenral} reduces to
\begin{equation}\label{Qdynleft}
\frac{\partial f(m,t)}{\partial t} = \Gc \overline m \frac{\partial}{\partial m} \left[\frac{m-\overline m}{w}\, f(m,t) + \frac{\partial f(m,t)}{\partial m} \right] \:,
\end{equation} 
where $\overline m  = w-q$ is the mean rung number. 
Dynamics of the correlator $g^{(2)}(t)$ according to Eq.~\eqref{eq:g2_eq} is given by
\begin{align}
&g^{(2)}(t) = \frac{2}{\Nc^2} \int_0^{\infty} m\, \chi(m,t)\, dm \:,
\end{align}
where $\chi(m,t)$ satisfies dynamic equation~\eqref{Qdynleft} with initial condition $\chi(m,0) = m\, f_0(m)$. 
This initial condition can be presented as a sum of two contributions: $\overline m f^{(0)}(m)$ and  $(m-\overline m)\, f^{(0)}(m)$. The first one does not evolve with time and provides the correct limit $g^{(2)}(t) \to 1$ at $t \to \infty$, while the latter turns out to be the eigenfunction of the right-hand side of Eq.~\eqref{Qdynleft} with the eigenvalue $-\Gc \overline m/w$. Hence, this eigenvalue describes the decay of the correlator $g^{(2)}(t)$ to unity, given in Eq.~\eqref{Qg2t} and upper part of Eq.~\eqref{Qtaug2}.

In the opposite case of the pumping rate larger than the critical $w-q \gg \sqrt{2w}$ ($W>W^*$) the stationary distribution function  Eq.~\eqref{Qf0} reduces to thermal one,
\begin{equation}\label{Qf0right}
f^{(0)}(m) = \frac{1}{2 \overline m}\, \e^{-m/\overline{m}}
 \:,
\end{equation}
where now the mean rung number is given by  $\overline m=q/(q-w)$.
Using distribution function Eq.~\eqref{Qf0right} we calculate the analytic expressions for $\Nx$ and $\Nc$ presented in the lower parts of Eqs.~\eqref{QNcNxg20}. However, in order to get a correction to $g^{(2)}(0)=2$ it is indispensable  to take into account a deviation of the static distribution from thermal. This can be done by introducing a factor $1 + ( \overline m^2 - m^2/2)\,w/q^2 $ into  Eq.~\eqref{Qf0right}.
Dynamic equation for the pumping higher than critical reduces to
\begin{equation}\label{Qdynright}
\frac{\partial f(m,t)}{\partial t} = \frac{\partial}{\partial m} \left[ \Gc m \left(\frac{f(m,t)}{\overline m} + \frac{\partial f(m,t)}{\partial m} \right) \right] \:.
\end{equation}
One can easily check that $(m-\overline m)\,f^{(0)}(m)$ is again the eigenfunction of the right-hand side of Eq.~\eqref{Qdynright}. Thus, the decay of the correlator $g^{(2)}(t)$ is governed by the corresponding rate $\Gc/\overline m$, which leads to Eq.~\eqref{Qg2t} and the lower part of Eq.~\eqref{Qtaug2}.

%

\end{document}